\documentclass[useAMS,usenatbib]{mn2e}
\usepackage{psfig}   
\usepackage{graphicx}
\usepackage{subfigure}

\newcommand{\apj}{ApJ}                                         

\newcommand{\apjl}{ApJL}
\newcommand{\aap}{A{\&}A}

\newcommand{\mnras}{MNRAS}
\newcommand{\aj}{AJ}

\newcommand{\Msun}{\ensuremath{\rm{M}_{\odot}}}

\title[Using the MST to trace mass segregation]{Using the minimum spanning tree to trace mass segregation}

\author[R.~J.~Allison et al.]{Richard~J.~Allison$^1$\thanks{E-mail:
  r.allison@sheffield.ac.uk}, Simon P.~Goodwin$^1$,
  Richard~J.~Parker$^1$, \newauthor Simon~F.~Portegies~Zwart$^{2,3}$,
  Richard~de~Grijs$^{1,4}$ and
  M.~B.~N.~Kouwenhoven$^1$\vspace*{0.1cm}\\ $^1$ Department of Physics
  and Astronomy, University of Sheffield, Sheffield, S3 7RH, UK\\ $^2$
  Section Computational Science, University of Amsterdam, Amsterdam,
  The Netherlands \\ $^3$ Astronomical Institute ``Anton Pannekoek'',
  University of Amsterdam, Amsterdam, The Netherlands \\ $^4$
  NAOC-CAS, Beijing, China}

\begin{document}

\date{}
                             
\pagerange{\pageref{firstpage}--\pageref{lastpage}} \pubyear{2008}

\maketitle

\label{firstpage}

\begin{abstract}
We present a new method to detect and quantify mass segregation in
star clusters.  It compares the minimum spanning tree (MST) of massive
stars with that of random stars.  If mass segregation is present, the
MST length of the most massive stars will be shorter than that of
random stars.  This difference can be quantified (with an associated
significance) to measure the degree of mass segregation.  We test the
method on simulated clusters in both 2D and 3D and show that the
method works as expected.

\noindent We apply the method to the Orion Nebula Cluster (ONC) and
show that the method is able to detect the mass segregation in the
Trapezium with a `mass segregation ratio' $\Lambda_{MSR}=8.0 \pm 3.5$
(where $\Lambda_{MSR}=1$ is no mass segregation) down to 16 \Msun,
and also that the ONC is mass segregated at a lower level ($\sim 2.0
\pm 0.5$) down to 5 \Msun.  Below 5 \Msun we find no evidence for
any further mass segregation in the ONC.
\end{abstract}


\section{Introduction}

It is thought that many young star clusters show `mass segregation' --
where the massive stars are more centrally concentrated than
lower-mass stars (see Section~2 for a review of current methods for
detecting mass segregation).  

There are two possible origins of mass segregation in star clusters.
It may be dynamical, in that mass segregation evolves within an
initially non-mass segregated cluster due to dynamical effects
\citep{chandrasekhar42,spitzer69}.  Or it may be primordial, where
mass segregation is an outcome of the star formation process
\citep{murray96,bonnell06}.  Information about the presence and degree
of mass segregation thus provides strong constraints on star cluster
formation and evolution.  However, there are currently no good methods
by which mass segregation can be determined and quantified absolutely.
This seriously limits the interpretation of observations of mass
segregation, and especially our ability to compare different clusters.

In this paper we present a new method, based on the minimum spanning
tree, which allows us to quantify the degree of mass segregation in a
cluster, and examine how mass segregation changes with stellar mass.
In Section~2 we review the limitations of current methods of
determining mass segregation.  In Sections~3 and~4 we introduce our
new method and test it against a number of artificial clusters and the
ONC.  In Section~5 we discuss the method's limitations and some ways
in which it might be improved, and we summarise our results in Section~6.

\section{Mass segregation}
\label{sec:ms}

\subsection{What is mass segregation?}
\label{ssec:whatms}

In this paper we will take as a working definition of mass segregation
that {\em the most massive stars are not distributed
  in the same way as other stars}.  In particular, that they
have a more concentrated distribution.

This is not the only possible definition of mass segregation.  It
would be possible to define mass segregation as a significant
difference in the mass functions\footnote{We deliberately distinguish
between the present-day mass function and the {\em initial} mass
function.} (MFs) at different positions in a cluster. Indeed, this is
often the way mass segregation is currently defined (see
Section~\ref{sec:currentmethods}).  However, we feel such a definition
is problematic as it raises the question of how many stars need to be
included in order to properly sample the MF and therefore tell if two
MFs are different.  Also, how much do the mass ranges sampled need to
overlap in order to differentiate MF slopes? Both of these questions
are difficult to answer without assuming a MF a priori.

\subsection{Current methods for detecting mass segregation}
\label{sec:currentmethods}

The current methods can be loosely separated into two groups,
those that involve fitting a density profile and characteristic radii
to various mass ranges; and those that trace the variation of the MF 
with radius.

None of the current methods to examine mass segregation give a 
model independent, quantitative measure of the amount of mass segregation 
in a cluster (and most are not quantitative at all).  The current 
methods of determining the presence of
mass segregation also rely on a determination of the centre of the
cluster, which in itself may be very difficult to define. 

Density profiles are usually fitted with a model profile (e.g. a King
(1966) model) or as a cumulative distribution function for different
mass `bins'. This allows the variation in distribution to be easily
seen, and the `degree' of mass segregation can be measured by the
change of the slope of the distribution function. From these
distributions characteristic radii for each mass bin (e.g. core
radius) can be found. This method has been used in many studies
\citep{hillenbrand97,pinfield98,raboud99,adams01,
  littlefair03,sharma08}.  However, the accuracy of this method has
been brought into question by \citet{gouliermis04}, who show that this
method is highly dependent on the number of mass bins, the size of the
bins, and on the models used to fit the density
profile. \citet{converse08} use a method which is similar to comparing
cumulative distributions. They study the amount of mass inside a
projected radius and obtain a quantifiable measure of the degree of
mass segregation from the area between this curve and a curve for a
cluster with no mass segregation.  However, this relies on having a
model for a non-mass segregated cluster. \citet{gouliermis08} use the
variation of the `Spitzer radius' (the
    Spitzer radius is the r.m.s. distance of 
stars in a cluster around the centre of mass) with 
luminosity to determine the presence of mass segregation.

A variation of the MF with radius will show the presence of mass
segregation as a shallowing of the slope of the MF with radius
\citep{degrijs02a,degrijs02b,degrijs02c,gouliermis04,
sabbi08,kumar08,harayama08}. Mass functions are calculated for various
annuli, such that in a mass segregated cluster the radii closest to
the centre of the cluster will contain the most massive stars, and so
they will have shallower slopes than those at larger radii. This
method also suffers from a strong dependence on the choice of annuli
and mass bins for the MF \citep{ascenso08}.  \citet{gouliermis04} find
that a unique set of annuli to use for comparison is not simple to
choose because of the variation of the shape and slope of the MF as
the choice of bin size changes for individual clusters.  A particular
problem is that low-mass stars are often difficult to detect in the
central regions leading to relatively small regions of overlap in the
MFs at different radii (which are the slopes to be compared).  A
further complication is introduced by the fact that the cluster centre
must be determined; it is generally chosen as where the luminosity is
highest - i.e. where the massive stars are.

\section{Minimum spanning tree method}
\label{sec:mst}

In this section we detail our new method for defining and quantifying
mass segregation using a minimum spanning tree (MST).

\subsection{The MST}

The MST of a sample of points is the path connecting all points in a
sample with the shortest possible pathlength, which contains no closed
loops \citep[see, e.g.][]{prim57}.  The length of the MST is unique,
but the exact path might differ in some special circumstances.  For
example, an equilateral triangle of stars named 1, 2, and 3 has two
possible MSTs -- one connecting stars 1 \& 2 and 2 \& 3, and one
connecting stars 1 \& 2 and 1 \& 3.  But the lengths of both MSTs are
identical.

We have used the algorithm of \citet{prim57} (see also
\citet{cartwright04}) to construct our MSTs.  We construct an
ordered list of the separations between all possible pairs of stars.
Stars are then connected together in `nodes' starting with the
shortest separations and proceeding through the list in order of
increasing separation, forming new nodes as long as the formation of
the node does not create a closed loop.

\subsection{Quantifying mass segregation}

Following our definition of mass segregation (see
Section~\ref{ssec:whatms}) we can detect and quantify mass segregation
by comparing the typical MST of cluster stars with the MST of the most
massive stars.

The method is based on finding the MST of the $N$ most massive stars
and comparing this to the MST of sets of $N$ random stars in the
cluster.  If the length of the MST of the most massive stars is
significantly shorter than the average length of the MSTs of the
random stars, then the massive stars have a different, and more
concentrated, distribution -- hence the cluster is mass segregated.
The ratio of the average random MST length to the massive star MST
length gives a quantitative measure of the degree of mass segregation
with an associated error (i.e. how likely it is that the massive star
MST is the same as that of a random set of stars).

The algorithm proceeds in the following way:

\noindent{\em 1. Determine the length of the MST of the $N_{\rm MST}$
most massive stars; $l_{\rm massive}$}.  The MST of a subset of the $N_{\rm
MST}$ most massive (or what ever subset is of interest) stars is
constructed.  This MST has length $l_{\rm massive}$.

\noindent{\em 2. Determine the average length of the MST of sets of
$N_{\rm MST}$ random stars; $\langle l_{\rm norm}\rangle$}.  Sets of
$N_{\rm MST}$ random stars are constructed, and the average length,
$\langle l_{\rm norm}\rangle$ of their MSTs is determined.  There is an
dispersion associated with the average length of the random MSTs which is
roughly gaussian and so can be quantified by the standard deviation of
the lengths $\langle l_{\rm norm}\rangle \pm \sigma_{\rm norm}$.

Numerical experiments have shown that $50$ random sets 
is sufficient to obtain a good estimate of the errors. However using
hundreds of random sets tends to result in smoother trends.  We
suggest using $500$ -- $1000$ random sets for low $N_{\rm MST}$ and at
least $50$ for high $N_{\rm MST}$.

Note that the error $\sigma_{\rm norm}$ on $\langle l_{\rm norm}\rangle$
is always large for small $N_{\rm MST}$ due to stochastic effects when
randomly choosing a small number of stars (see Section~\ref{sec:msclusters}).

\noindent{\em 3. Determine with what statistical significance $l_{\rm
  massive}$ differs from $\langle l_{\rm norm}\rangle$}.  We define the
  `mass segregation ratio' ($\Lambda_{MSR}$) as the ratio between the average
  random path length and that of the massive stars:
%----eq:msr----------
\begin{equation}
\Lambda_{MSR}=\frac{\langle l_{\rm norm}\rangle}{l_{\rm massive}} \pm \frac{\sigma_{\rm norm}}{l_{\rm massive}}
\label{eq:msr}
\end{equation}
%----eq:msr-
where an $\Lambda_{MSR}$ of $\sim 1$ shows that the massive stars are distributed
in the same way as all other stars, an $\Lambda_{MSR}$ significantly $>1$ indicates
mass segregation, and an $\Lambda_{MSR}$ significantly $<1$ indicates inverse-mass
segregation (i.e. the massive stars are more widely spaced than other
stars).  The more (significantly) different the $\Lambda_{MSR}$ is from unity, the
more extreme is the degree of (inverse-)mass segregation.

\noindent{\em 4. Repeat the above steps for different values of
  $N_{\rm MST}$ to determine at what masses the cluster is segregated,
  and to what degree at each mass.}  Clearly, the choice of $N_{\rm
  MST}$ is arbitrary, so the $\Lambda_{MSR}$ needs to be determined for many
  values of $N_{\rm MST}$ to gain information on how the degree of
  mass segregation changes with mass.

This method has three major advantages over current methods.  Firstly,
it gives a quantitative measure of mass segregation with an associated
significance, allowing different clusters to be directly compared.
Secondly, it does not rely on defining a cluster centre or any special
location in a cluster (allowing it to be applied to highly
substructured regions as well).  Thirdly, it is applicable not just to
the most massive stars, but to any subset of stars (or brown dwarfs)
that one might think has a different distribution to the `norm'.

It should be noted that whilst we are applying our method to simulated
data and therefore know the mass of each star, an absolute
determination of the mass is not required, just a measure of the
relative masses (i.e. luminosity).

\section{The method in action}
\label{sec:applications}

\subsection{A non-mass segregated cluster}

We tested the method on a large sample of Plummer spheres
\citep{plummer11} consisting of 1000 single stars with masses sampled
from a Kroupa initial mass function \citep{kroupa02}.  The masses are
assigned randomly to stars so these clusters should not be mass
segregated (except by chance).

Tables~\ref{tab:randplummer3d} and~\ref{tab:randplummer2d} show the
fractions of random clusters that are found at $1 \sigma$ significance
to be either inverse-mass segregated (Inverse-MS: $\Lambda_{MSR}<1$), mass segregated
(MS: $\Lambda_{MSR}>1$) or that showed no mass segregation (No MS: $\Lambda_{MSR}$ $\sim 1$)
for various numbers of members in the MST
($N_{\mathrm{MST}}$). Table~\ref{tab:randplummer3d} uses the 3D
positions of the stars, Table~\ref{tab:randplummer2d} is for a 2D
projection as would be observed.  In both the 2D and 3D clusters we
find that false positives and negatives (i.e. $\Lambda_{MSR}$ $>1 \sigma$ from
unity) occur with equal frequency around $1/6$th of the time -- exactly
what would be expected if the error is Gaussian.

%----tab:randplummer3d----------
\begin{center}
\begin{table}
  \begin{tabular}{|r|r|r|r|r|}
\hline
        & \multicolumn{4}{|c|}{$N_{\mathrm{MST}}$} \\
        & 5      & 10   & 100  & 200        \\ \hline
Inverse-MS & 14.1\% & 15.5 & 17.1 & 15.2       \\
MS      & 12.6\% & 14.1 & 17.0 & 17.2       \\
No MS   & 73.3\% & 70.4 & 65.9 & 67.6       \\ \hline
  \end{tabular}
\caption{The percentage of Plummer spheres which show evidence of
inverse-mass segregation (Inverse-MS), mass segregation (MS), and no mass
segregation (No MS) for various numbers of members in the MST, 
$N_{\mathrm{MST}}$, in 3 dimensions.
\label{tab:randplummer3d}}
\end{table}
\end{center}
%----tab:randplummer3d----------
%----tab:randplummer2d----------
\begin{center}
\begin{table}
  \begin{tabular}{|r|r|r|r|r|}
\hline
        & \multicolumn{4}{|c|}{$N_{\mathrm{MST}}$} \\
        & 5      & 10   & 100  & 200  \\ \hline
Inverse-MS & 13.1\% & 14.9 & 17.2 & 17.1 \\
MS      & 10.4\% & 12.2 & 17.0 & 19.1 \\
No MS   & 76.5\% & 72.9 & 65.8 & 63.8 \\ \hline
  \end{tabular}
\caption{As Table~\ref{tab:randplummer3d} but for clusters projected
  into 2 dimensions as would be observed.
\label{tab:randplummer2d}}
\end{table}
\end{center}
%----tab:randplummer2d----------

\subsection{Mass segregated clusters}
\label{sec:msclusters}

We then test the method with mass segregated clusters. These are
created by first producing Plummer spheres of 1000 stars in the same
way as before.  The positions of the most massive $x$ per cent of the
stars are then swapped with randomly chosen stars positioned inside
the $x$ per cent number radius, so only the most massive stars are
inside this inner radius. The routine creates clusters in which the
most massive stars are centrally concentrated, and so are mass
segregated.  This is not necessarily a realistic configuration as
Nature might produce, but it is one in which we can easily control the
degree of mass segregation.

Each cluster contains 1000 stars.  Therefore, a 5 per cent MS places
the 50 most massive stars within the inner 5 per cent number radius.
For 10 per cent MS, it is the 100 most massive stars, and so on. We
vary the radii in which the most massive stars are placed from the 5
per cent number radius to the 80 per cent number radius.  Note that at
large values the cluster is more like one in which the low-mass stars
have been concentrated at the outskirts rather than one in which the
high-mass stars are centrally concentrated.

Tables~\ref{tab:msplummer3d} and~\ref{tab:msplummer2d} show the
percentage of clusters which the method finds to have mass segregation
at $> 1 \sigma$ significance for different numbers of stars in the MST
($N_{\rm MST}$).  In brackets after the percentage is the typical
significance of the mass segregation. As before,
Table~\ref{tab:msplummer3d} uses the 3D positions of the stars, while
in Table~\ref{tab:msplummer2d} a 2D projection is used.
 
%----tab:msplummer3d----------
\begin{table}
  \begin{tabular}{|c|c|c|c|c|c|}
    \hline
\%MS & \multicolumn{5}{|c|}{$N_{\mathrm{MST}}$}           \\
     & 5       & 10      & 100     & 200      & 500       \\ \hline
5    & 100 (2) & 100 (3) & 100 (3) & 90 (2)   & 40 (1)    \\
10   & 100 (2) & 100 (3) & 100 (7) & 100 (4)  & 70 (1)    \\
20   & 100 (2) & 100 (3) & 100 (7) & 100 (10) & 100 (3)   \\ 
50   & 80 (2)  & 100 (2) & 100 (5) & 100 (8)  & 100 (15)  \\
80   & 30 (1)  & 80 (1)  & 100 (4) & 100 (6)  & 100 (10)  \\ \hline
  \end{tabular}
\caption{The frequency of a positive detection of mass segregation for
various $N_{\rm MST}$ and various degrees of mass segregation (\%MS)
in a 3D cluster.  The numbers in parentheses denote the typical
significance (in sigma) at which mass segregation is detected.
\label{tab:msplummer3d}}
\end{table}
%----tab:msplummer3d----------
%----tab:msplummer2d----------
\begin{table}
  \begin{tabular}{|c|c|c|c|c|c|}
    \hline
\%MS  & \multicolumn{5}{|c|}{$N_{\mathrm{MST}}$}          \\
      & 5       & 10      & 100     & 200      & 500      \\ \hline
5     & 100 (2) & 100 (2) & 100 (3) & 70 (2)   & 30 (2)   \\
10    & 100 (2) & 100 (2) & 100 (6) & 100 (2)  & 50 (2)   \\
20    & 100 (2) & 100 (2) & 100 (6) & 100 (9)  & 80 (2)   \\ 
50    & 70 (1)  & 100 (2) & 100 (5) & 100 (7)  & 100 (12) \\
80    & 0 (-)   & 60 (1)  & 100 (3) & 100 (6)  & 100 (10) \\ \hline
  \end{tabular}
\caption{As Table~\ref{tab:msplummer3d}, but for a 2D projection as
  would be observed.
\label{tab:msplummer2d}}
\end{table}
%----tab:msplummer2d----------

It is clear from Tables~\ref{tab:msplummer3d}
and~\ref{tab:msplummer2d} that the ability of the method to detect
mass segregation depends on a combination of the level of mass
segregation and the number of stars in the MST. In the first column,
when $N_{\rm MST} = 5$ the method is able to detect mass segregation
at roughly $2 \sigma$ significance when the \%MS is low.  As it only
uses the 5 most massive stars to look for mass segregation, it is not
good at finding mass segregation when the \%MS involves hundreds of
stars.

In contrast, when $N_{\rm MST} = 500$ (i.e. half of the stars in the
cluster), the method is poor at spotting when the mass segregation
only involves a few stars.  However, it becomes extremely good when
finding mass segregation involving many hundreds of stars.

Looking across Tables~\ref{tab:msplummer3d} and~\ref{tab:msplummer2d},
the highest significances for finding mass segregation (up to 10 or
15$\sigma$ results) occur when $N_{\rm MST}$ is equal to the number of
stars that have been mass segregated.  The best results are for large
$N_{\rm MST}$ as the variance in $\langle l_{\rm norm}\rangle$ is
smallest when $N_{\rm MST}$ is large.  Note that, as mentioned above,
when $N_{\rm MST}$ is small the variance in $\langle l_{\rm
norm}\rangle$ is large due to stochastic effects.

Therefore, it is vital to vary $N_{\rm MST}$ in order to examine mass
segregation.  However, this is an important diagnostic tool, as
variations of $\Lambda_{MSR}$ and its significance with $N_{\rm MST}$ contain
information about how, and down to which mass mass segregation is
found.  

%----fig:msplummernmst----------
\begin{figure}
  \begin{center}
    \setlength{\subfigcapskip}{10pt}
    \includegraphics[scale=0.7]{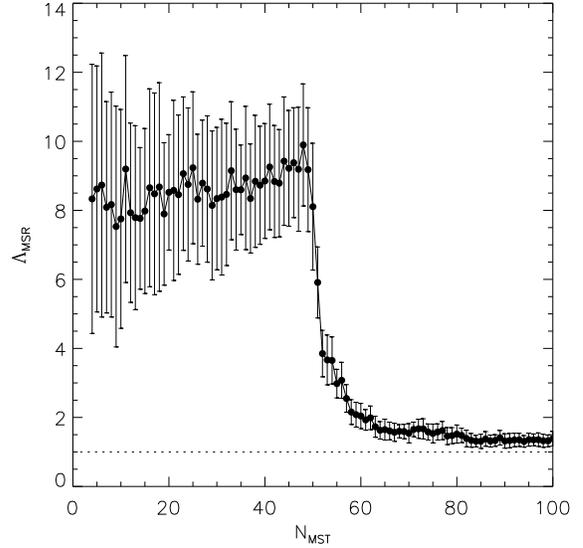}
%\hspace*{0.6cm}
  \end{center}
  \caption[bf]{The evolution of $\Lambda_{MSR}$ with $N_{\rm MST}$ for
  a Plummer sphere with an initial 5 percent MS. The dashed line
  indicates $\Lambda_{MSR}$=1 i.e. No MS.}
  \label{fig:nmstmspl05}
\end{figure}
%----fig:msplummernmst----------

In Fig.~\ref{fig:nmstmspl05} we show the variation of the $\Lambda_{MSR}$ with
$N_{\rm MST}$ for a 5 per cent level of mass segregation (i.e. the 50
most massive stars are segregated).  The $\Lambda_{MSR}$ is around $9$ for $N_{\rm
  MST} < 50$, and then drops sharply toward unity when $N_{\rm MST} >
50$.  This shows two features of mass segregation that can be
extracted using this method.

Firstly, it shows that only the $50$ most massive stars are mass
segregated (as described above).  However, it also shows that the $50$
most massive stars are {\em equally} mass segregated.  That is, the
method places the $50$ most massive stars at the centre of the cluster
but does not order them in any specific way within the centre.

%----fig:2msplnmst----------
\begin{figure}
  \begin{center}
    \setlength{\subfigcapskip}{10pt}
    \includegraphics[scale=0.7]{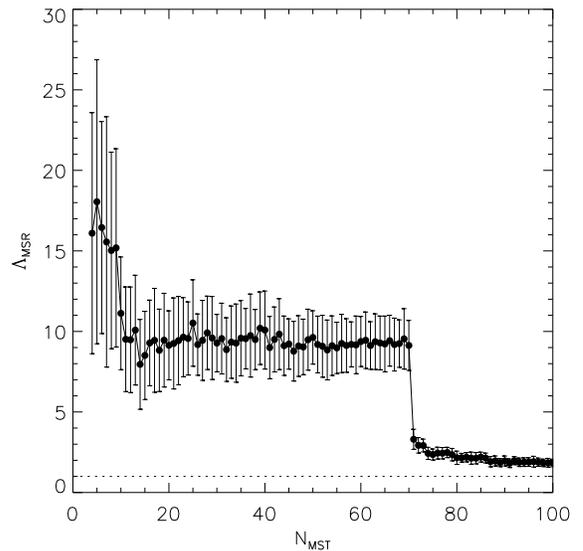}
%\hspace*{0.6cm}
  \end{center}
  \caption[bf]{The evolution of $\Lambda_{MSR}$ with $N_{\rm MST}$ for
  a Plummer sphere with an initial 1 per cent and 7 per cent MS. The
  dashed line indicates $\Lambda_{MSR}$=1 i.e. No MS.}
  \label{fig:nmst.2mspl}
\end{figure}
%----fig:2msplnmst----------

To illustrate this we show in Fig.~\ref{fig:nmst.2mspl} the variation
of the $\Lambda_{MSR}$ with $N_{\rm MST}$ for a cluster in which the $10$ most
massive stars are placed in the inner $1$ per cent radius, and
the $11$th to $70$th most massive stars in the $1$ -- $7$ per cent
radius.  The method clearly detects that there are two levels of mass
segregation -- one involving the 10 most massive stars, and one which
involves the 70 most massive stars.

\subsection{A complex mass segregated cluster}

%----fig:complexnmst----------  f1000.a1.01 SS#27 2D MST
\begin{figure}
  \begin{center}
    \setlength{\subfigcapskip}{10pt}
    \includegraphics[scale=0.4,angle=270]{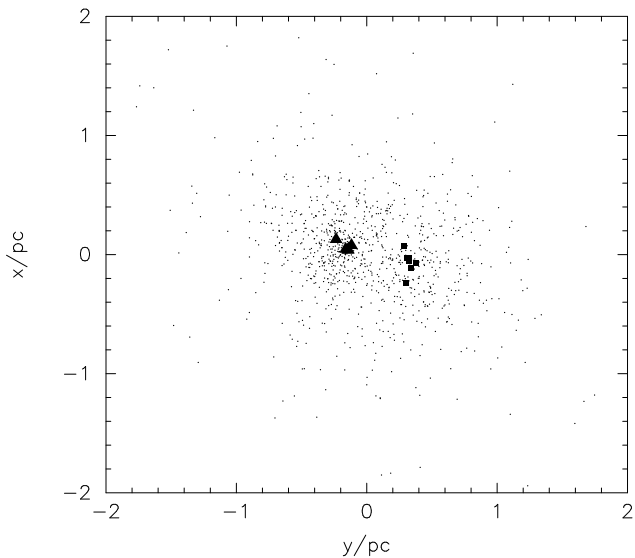}
%\hspace*{0.6cm}
  \end{center}
  \caption[bf]{The distribution of stars in a cluster evolved from
  cold, fractal initial conditions.  The triangles show the positions
  of the five most massive stars, the squares the positions of the 6th
  to 12th most massive stars.}
  \label{fig:cluster}
\end{figure}
%----fig:complexnmst----------

We also apply the method to a more realistic and complex cluster.  In
Fig.~\ref{fig:cluster} we show a complex 1000-body
cluster\footnote{The cluster has been evolved from a cold fractal
distribution \citep[see,][]{goodwin04}; we will discuss simulations
like this in detail in a later paper, for now the cluster is merely
used as an illustration of the method.}.  The triangles show the
positions of the 5 most massive stars and the squares the positions of
the 6th to 12th most massive stars (the reason for these choices will
be explained below).

Interestingly, the cluster has collapsed into a configuration in which
the 5 most massive stars form a dense Trapezium-like cluster to one
side, and the 6th to 12th most massive stars have formed a separate
`clump' on the other side.  For this reason the cluster has no
well-defined centre which serves to illustrate the strength of the MST
method in dealing with unusual and clumpy clusters.  

\begin{figure}
  \begin{center}
    \setlength{\subfigcapskip}{10pt}
    \includegraphics[scale=0.7]{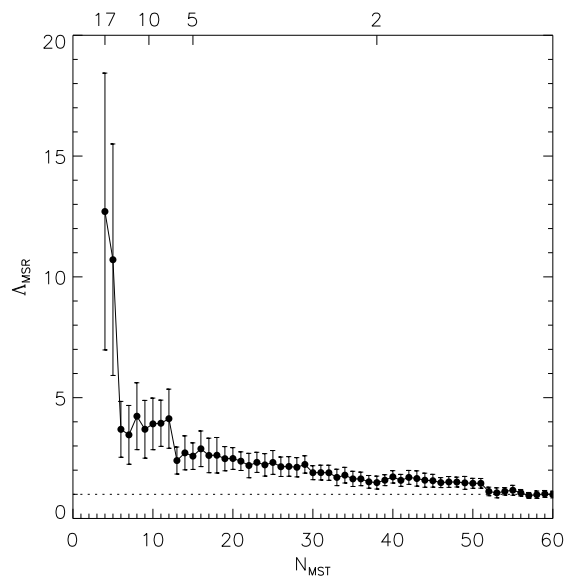}
%\hspace*{0.6cm}
  \end{center}
  \caption[bf]{The evolution of the 2D $\Lambda_{MSR}$ with $N_{\rm MST}$ for 
the cluster illustrated in fig.~\ref{fig:cluster}.  The dashed 
line indicates an $\Lambda_{MSR}$ of unity, i.e. no mass segregation.}
  \label{fig:nmstcomplex}
\end{figure}

In Fig.~\ref{fig:nmstcomplex} we show the change of the 2D $\Lambda_{MSR}$ with
$N_{\rm MST}$ for the cluster illustrated in Fig.~\ref{fig:cluster}. 
Fig.~\ref{fig:nmstcomplex} shows that
in this cluster there are 3 `levels' of mass segregation.
\textit{Firstly}, there is the presence of a Trapezium-like system
formed from the five most massive stars (with an $\Lambda_{MSR}$ of
around 12).  \textit{Secondly}, there is another system, containing 
the 6th to 12th most massive stars, which shows significantly less 
mass segregation than the Trapezium-like
system (an $\Lambda_{MSR}$ of around 4).  \textit{Thirdly}, the 50 most massive
stars in the cluster show a steadily decreasing degree of mass
segregation.  Beyond the 50th most massive star the cluster shows no
evidence for mass segregation at all.

We note that the second clump was not obvious in our initial
`eyeballing' of the simulation.  It only became obvious when we
plotted the positions of the 6th to 12th most massive stars after the
MST method alerted us to there being something special about these
stars.

%----fig:complexnmst----------
%----fig:complexnmst----------

%----fig:ONCnmst----------
\begin{figure}
  \begin{center}
    \setlength{\subfigcapskip}{10pt}
    \includegraphics[scale=0.7]{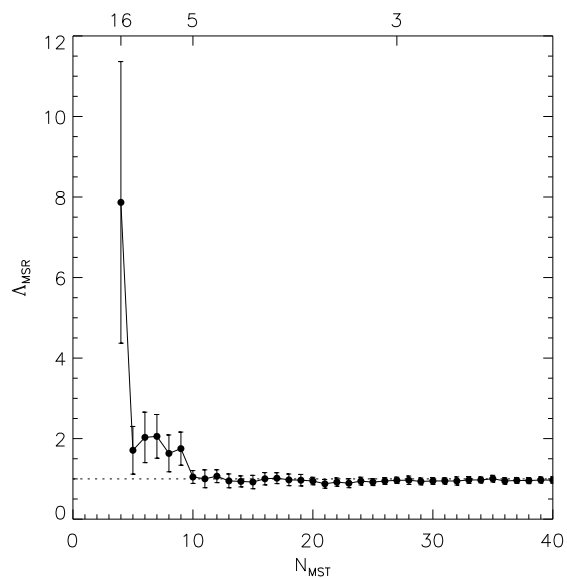}
%\hspace*{0.6cm}
  \end{center}
  \caption[bf]{The evolution of $\Lambda_{MSR}$ with $N_{\rm MST}$ for the
  ONC. The dashed line indicates an $\Lambda_{MSR}$ of unity, i.e. no mass
  segregation.}
  \label{fig:nmstONC}
\end{figure}
%----fig:nmstONC----------

\subsection{The Orion Nebula Cluster}

Finally, we apply the method to real data, specifically the Orion
Nebula Cluster (ONC) data of \citet{hillenbrand97}.  We use the 900
stars for which \citet{hillenbrand97} provide masses.  We note that
this is not an ideal dataset as many (presumably low-mass) stars lack
masses. However it serves to illustrate the method.

\citet{hillenbrand98} show, using cumulative distribution functions
and mean stellar mass as a function of radius, that the ONC is mass
segregated. They find evidence for mass segregation in the ONC for
stars more massive than 5 \Msun within 1 pc, with less compelling
evidence beyond this radius.

Figure~\ref{fig:nmstONC} shows the change of $\Lambda_{MSR}$ with
$N_{\rm MST}$ for the ONC.  The MST method clearly picks out the mass
segregation of the Trapezium system at $N_{\rm MST} = 4$ with
$\Lambda_{MSR}=8.0 \pm 3.5$.  However, the MST method also shows that
there appears to be a secondary level of mass segregation involving
the 9 most massive stars ($> 5$ \Msun) in the ONC in agreement with
\citet{hillenbrand98}.  Whilst the significance of the $\Lambda_{MSR}$
is low (around $2.0 \pm 0.5$), there is a clear trend in that each of
the 5th to 9th stars show the same increased level of mass
segregation.  However, unlike \citet{hillenbrand98} we find no
evidence of mass segregatation below 5 \Msun.

\section{Further applications and future work}

In future papers we will apply our new method to more observational
and theoretical data to look for
mass segregation, and also to examine if low-mass stars have
different distributions to brown dwarfs and/or high mass stars.

{\em Binaries}.  In our tests we have not included binary stars, and
they will have an effect on the lengths of MSTs.  If a binary is
resolved, then it is very likely that the two components will be
linked as a node (if this is not the case it would be unclear that the
system really was a binary).  Indeed, triples and quadruples will also
be linked as a subset within the MST.  This raises the possibility 
that MSTs could be very useful in locating binary and multiple 
systems by looking for short links within the MST.  However, as noted
by Cartwright \& Whitworth (2005) care must be taken in dealing with
binaries.  

{\em Incompleteness}.  In many clusters there is significant
incompleteness, especially in regions around the most massive stars
where low-mass stars cannot be observed (if they are present).  This
presents a significant problem as it is impossible to know if there
are many low-mass stars in the `central' regions.  If there are, then
an average MST length would be short. If not, it would be long (this
is exactly the same problem faced by other methods).  We will explore
this problem in a future paper by creating and analysing synthetic
observational datasets.

\section{Summary}
\label{sec:conclusion}

We have outlined a new method of determining and quantifying mass
segregation in a cluster by using a minimum spanning tree (MST). The
algorithm proceeds in the following way: 

{\em 1. Determine the length of the MST of the $N_{\rm MST}$ most
massive stars; $l_{\rm massive}$}

{\em 2. Determine the average length of the MST of sets of $N_{\rm
MST}$ random stars; $\langle l_{\rm norm}\rangle$}

{\em 3. Determine with what statistical significance $l_{\rm massive}$
differs from $\langle l_{\rm norm}\rangle$.}

%----eq:msr----------
\begin{displaymath}
\Lambda_{MSR}=\frac{\langle l_{\rm norm}\rangle}{l_{\rm massive}} \pm \frac{\sigma_{\rm norm}}{l_{\rm massive}}
\end{displaymath}
%----eq:msr-

{\em 4. Repeat the above steps for different values of $N_{\rm MST}$
  to determine at what masses the cluster is segregated, and to what
  degree at each mass.}

By examining the difference between the MST of a subset of (massive)
stars and an equal number of random stars it is possible to quantify
the level of (inverse-)mass segregation in a cluster.  Tests on artificial
clusters show that the method behaves as expected and can identify
mass segregation.

We apply the method to the Orion Nebula Cluster (ONC) and find that
the Trapezium is mass segregated with an $\Lambda_{MSR}$ of $8.0 \pm
3.5$.  We also find that the 5th to 12th most massive stars down to
around 5 \Msun also show evidence of being mass segregated with an
$\Lambda_{MSR}$ of $2.0 \pm 0.5$.  Below 5 \Msun we find no evidence
for mass segregation.

\section*{Acknowledgements}

We thank Stuart Littlefair for useful discussions. RJA and RJP
acknowledge financial support from STFC. MBNK was supported by
PPARC/STFC under grant number PP/D002036/1. SPZ is grateful for the
support of the Netherlands Advanced School in Astrophysics (NOVA), the
LKBF and the Netherlands Organization for Scientific Research (NWO).We
acknowledge the support and hospitality of the International Space
Science Institute in Bern, Switzerland where part of this work was
done as part of a International Team Programme.

%---------------------------------------------------------------------------

\label{lastpage}

%\newcommand{ \bibfont}{\small}
%\setlength{\bibsep}{0pt}
%\bibliography{allison_mst}
%\bibliographystyle{mn2e}

\end{document}